\documentclass[conference]{IEEEtran}
\IEEEoverridecommandlockouts
\usepackage{cite}
\usepackage{amsmath,amssymb,amsfonts}

\usepackage{graphics}
\usepackage{algorithmic}
\usepackage{graphicx}
\usepackage{placeins}
\usepackage{caption}
\usepackage[]{hyperref}
\usepackage{textcomp}

\usepackage{xcolor}
\def\BibTeX{{\rm B\kern-.075em{\sc i\kern-.055em b}\kern-.010em
    T\kern-.2em\lower.9ex\hbox{E}\kern-.150emX}}

\begin{document}

\title{A Systematic Literature Review On Privacy Of Deep Learning Systems \\ [2ex] \Large Vishal Jignesh Gandhi, Sanchit Shokeen, Saloni Koshti }

\author{University Of Guelph \\ 
\hrulefill}

\maketitle

\begin{abstract}

The last decade has seen a rise of Deep Learning with its applications ranging across diverse domains. But usually, the datasets used to drive these systems contain data which is highly confidential and sensitive. Though, Deep Learning models can be stolen, or reverse engineered, confidential training data can be inferred, and other privacy and security concerns have been identified. Therefore, these systems are highly prone to security attacks. This study highlights academic research that highlights the several types of security attacks and provides a comprehensive overview of the most widely used privacy-preserving solutions. This relevant systematic evaluation also illuminates potential future possibilities for study, instruction, and usage in the fields of privacy and deep learning. \\
\end{abstract}


\section{Introduction}

Deep Learning, which is simply a neural network with three or more layers, is a subset of machine learning. These neural networks make an effort to mimic how the human brain functions, however they fall far short of being able to match it, enabling it to "learn" from vast volumes of data. Due to the Internet's and traditional telecommunication networks' rapid development, a large number of terminal devices are connecting to the network and generating enormous amounts of data every day [\ref{rs:01}]. Many artificial intelligence (AI) apps and services are powered by Deep Learning, which enhances automation by carrying out mental and physical tasks without the need for human intervention. Deep Learning is the technology that powers both established and emerging technologies, like voice-activated TV remote controls, digital assistants, and credit card fraud detection. Furthermore, all these devices work on IoT (Internet of Things) [\ref{a1}]. IoT generates a huge amount of data, and directly integrating this data on one server will result in privacy leaks, especially for sensitive personal information. Distributed collaborative learning of a model is an alternative to centralized training that makes use of all these raw inputs [\ref{rs:08}].

Moreover, Deep Learning has recently demonstrated great performance in a variety of fields, including image recognition, pattern matching, and even cybersecurity. Many applications that we use every day to make decisions based on predictions use Deep Learning models, thus if these models were to mispredict the future due to malevolent internal or external factors, it might cause problems in our daily lives. Furthermore, the Deep Learning training models frequently contain sensitive user data, so those models shouldn't be exposed to security and privacy risks. For protection and confidentiality reasons, if data for training the model is restricted to a sole source it can severely impact the model accuracy and result in a sub-par model, because of overfitting [\ref{rs:03}, \ref{a2}]. Deep Learning and machine learning algorithms are still susceptible to many dangers and threats to their security. As a result, it is imperative to alert the sector to security dangers and apply appropriate Deep Learning countermeasure solutions [\ref{rs:02}]. 


\subsection{Prior research}
To the best of our knowledge, there are very limited reviews which summarise the distinct types of privacy attacks on Deep Learning architectures and the solutions usually deployed to protect our systems from such attacks. A study by Sheraz et al. [\ref{rs:03}], discusses the various threats related to privacy in a Deep Learning ecosystem and the potent strategies deployed to safeguard the systems from the distinct threats. It analyses the performance of these different techniques – Homomorphic Encryption, Differential Privacy and Secure Two-Party Computation, and talks about the various challenges in the domain.

In a review by Amine et al. [\ref{rs:04}], apart from weighing the different solutions on different benchmarks and standards, this study categorises the different strategies for maintaining privacy in Deep Learning, in a multi-level taxonomy. Key lessons from every privacy conserving strategy are also highlighted. It also provides insight into future research options and open challenges in the field.

In this work, Milad et al. [\ref{rs:05}] assess the various white-box inference attacks and their effect on privacy of Deep Learning systems is analysed in centralised and distributed learning. Security in Deep Learning is discussed briefly in two short surveys [\ref{rs:06},\ref{a4}] [\ref{rs:07}], with no dialogue on the future research opportunities and challenges in the domain. Collaborative learning is reviewed and deployed for preserving privacy in Deep Learning systems both during training and using, in the study by Zhang et al. [\ref{rs:06},\ref{a5}]. Privacy protection strategies along with the respective threats during both learning and the testing phases of Deep Learning are reviewed in the examination by Chang et al. [\ref{rs:07}, \ref{a3}].

On other hand, distributed Deep Learning training, we also have huge dataset parallelization strategies (such as model and data parallelization). To train deep neural networks distributable, data parallelization, which divides the input samples, is frequently utilized [\ref{rs:09},\ref{a6}].

However, it is challenging to apply these standard tactics when the data contain restrictions on sharing among multiple computing nodes and are sensitive in terms of security and privacy. In this paper, we present a strategy for training deep neural networks while maintaining dataset privacy by successively sharing models in cyclic sequence during training procedures [\ref{rs:09},\ref{a8}]. Additionally, there are two practical issues that can be resolved: the dataset imbalance issue and the effective use of idle bots' computational resources [\ref{rs:09},\ref{a7}].

\subsection{Research goal}
The analysis of prior research and its conclusions, as well as a summary of the research efforts in the area of Deep Learning system privacy, are the goals of this study. As stated in Table [\ref{tab:1}], we created three research questions to help us focus the effort

\begin{table}[hb]
\caption{}
    \centering
    \begin{tabular}{p{0.35\linewidth} | p{0.6\linewidth}}
      Research Question (RQ)  & Discussion \\ \hline
      
      RQ1: What are the different types of privacy attacks on a Deep Learning system? & Variety of attacks can compromise the functioning and performance of Deep Learning architecture. A discussion on the types of attacks can summarise the ways in which an attacker can exploit the weak links in the system. \\[0.5cm]
      
      RQ2: What are the various privacy-preserving solutions for the different attacks? & Different strategies are employed to enhance the privacy in Deep Learning networks. A mapping of the variety of techniques applied to protect the system from the various attacks will provide a clear understanding of the privacy preserving approaches.\\[0.5cm]
      
      RQ3: What are the methods by which we can preserve the privacy of users and reduce the communication burden? & A privacy-preserving framework used to enable various participants to distributively learn a model with a privacy protection promise in order to secure user privacy will be discussed. In addition, a novel gradient sparsification technique which lowers the communication costs for both upload and download is reviewed. \\
    \end{tabular}
    \label{tab:1}
\end{table}

\subsection{Contributions and layout}
For people interested in "privacy of Deep Learning systems" and cyber security, this SLR supports existing research and offers the following contributions to advance their work of Table [\ref{tab:2}]. \\

The format of this essay is as follows: The techniques used to choose the primary studies for analysis in a methodical manner are described in Section [\ref{sec:2}].
The results of all the primary research chosen are presented in Section [\ref{sec:3}]. The findings in relation to the earlier-presented study topics are discussed in Section [\ref{sec:4}]. The research is concluded in Section [\ref{sec:5}], which also makes some recommendations for more study. Furthermore, we also add a conclusion and future work in section [\ref{sec:6}].

\section{\textbf{Research Methodology}}\label{sec:2}
We attempted to progress through the planning, conducting, and reporting phases of the review in iterations to enable a full examination of the SLR in order to reach the goal of answering the research questions. 
\subsection{Selection of primary studies}
Passing keywords to a particular publication's or search engine's search function highlighted primary studies. The keywords were chosen to encourage the publication of research findings that would help answer the research questions. Only AND and OR were allowed to be used as Boolean operators. The query terms were:\\[0.5cm]
\textbf{("All Metadata": "deep learning" OR "All Metadata": "deep-learning" OR "All Metadata": "deeplearning") AND ("All Metadata": "privacy")}\\[0.5cm]

Total result - 1937 results\\
The platforms searched were: 
\begin{itemize}

\item Google Scholar - 47 

\item ACM library - 29  

\item IEEE Xplore Digital Library - 1861

\end{itemize}
Depending on the search platforms, the title, keywords, or abstract were used in the searches. On October 9, 2022, we did the searches and processed all studies that had been published up to that point. The inclusion/exclusion criteria, which will be provided in Section B, were used to filter the results from these searches. Snowballing iterations were performed both forward and backward until no further publications that met the inclusion criteria could be found. 

\subsection{Inclusion and exclusion criteria}
The research work chosen for this SLR should be focused mainly on the privacy of Deep Learning systems, different techniques presented on preserving the privacy of data transferred and stored while using various Deep Learning models and the papers discussing the privacy-preserving Deep Learning models. Google Scholar, ACM library and IEEE platforms were searched for shortlisting the papers and only the ones concentrated on the privacy preserving models of Deep Learning systems are selected. It is shown in Table [\ref{tab:2}].

\begin{table}[ht]
\caption{}
    \centering
    \begin{tabular}{p{0.35\linewidth} | p{0.6\linewidth}}
      Inclusion criteria   & Exclusion criteria  \\ \hline
      
      The papers mainly focusing on privacy of Deep Learning systems  & The papers selected must not research on the Deep Learning systems and its techniques instead the focus should be preserving the privacy of data used in Deep Learning models  \\[0.5cm]
      
      The papers must present different ways to preserve privacy in various fields of Deep Learning  & Websites and blogs on Deep Learning and its various models\\[1cm]
      
      Some papers were also selected that conducted a survey on Deep Learning privacy models and the challenges faced to preserve it & The papers must not be published before 2018. \\
    \end{tabular}
    \label{tab:2}
\end{table}

\subsection{Selection results}

The initial keyword searches on the chosen platforms turned up a total of 1937 studies. This was lowered to 735 when duplicate studies were eliminated. There were 114 publications left after the research were examined under the inclusion/exclusion criteria. After applying the inclusion/exclusion criteria again and reading all 114 papers, 54 papers were still present. We selected only those papers which are published in and after 2018. So, only 49 pieces of paper make up the entire SLR.

\subsection{Quality assessment}

An assessment of the quality of primary studies was made according
to the guidance set by Kitchenham and Charters[\ref{rs:25}]

To determine their effectiveness, five papers were chosen at random and put through the following quality assessment process.

Stage 1:\textbf{ Deep Learning privacy:} The paper should be focused on the usage of Deep Learning privacy and its application in that domain. \\

Stage 2:\textbf{ Context:} The research aims, and findings must be adequately contextualized. This will enable correct research interpretation. \\

Stage 3:\textbf{ Attacks on the Deep Learning system:} In order to accurately portray how the technology has been applied to a particular situation and help answer research questions, there must be enough information in the study, RQ 1 and RQ2. \\

Stage 4:\textbf{ Security context:} In order to answer RQ3, the privacy and security issue must be addressed, explained, and added to the article. \\

Stage 5:\textbf{ Deep Learning performance:} The privacy preserving solutions applied must not hamper the performance of the Deep Learning system by a significant margin. \\

Stage 6:\textbf{ Data acquisition:} Specifics regarding the data's collection, measurement, and reporting must be provided to assess correctness. \\

	All selected primary studies were then subjected to this checklist for quality assessment. According to the findings, 8 studies were eliminated from the SLR because they failed to satisfy one or more of the checklist requirements, as indicated in Table \ref{tab:3}.

 \begin{table}[ht]
\caption{Excluded studies}
    \centering
    \begin{tabular}{p{0.5\linewidth} | p{0.4\linewidth}}
      Checklist for the Criteria Stages  & Excluded Studies \\ \hline
      
     Stage 1: Deep Learning privacy & [[\ref{ps:33}]] [[\ref{ps:48}]] \\
      
      Stage 2: Context & [[\ref{ps:39}]]\\
      Stage 3: Attacks on the Deep Learning system  & [[\ref{ps:41}]]\\
      Stage 4: Security context  & [[\ref{ps:31}]] [[\ref{ps:27}]]\\
      Stage 5: Deep learning performance  & [[\ref{ps:23}]]\\
      Stage 6: Data acquisition & [[\ref{ps:37}]]\\
      
    \end{tabular}
    \label{tab:3}
\end{table}

\subsection{Data extraction}
Data was then taken from all papers that had passed the quality evaluation in order to evaluate the completeness of the data and verify the accuracy of the information included within the articles. Before being extended to cover the entire set of research that has passed the quality evaluation step, the data extraction technique was first tested on a preliminary five investigations. Each study's data was taken out, put into categories, and then entered into a spreadsheet. The following categories were applied to the data: \\

\textbf{Context data:} Information on the study's objectives. \\

\textbf{Qualitative data:} Results and recommendations offered by the authors.\\ 

\textbf{Quantitative data:} Data obtained from testing and research are then applied to the study. \\

\begin{figure}[ht]
\centerline{\includegraphics [width=70mm]{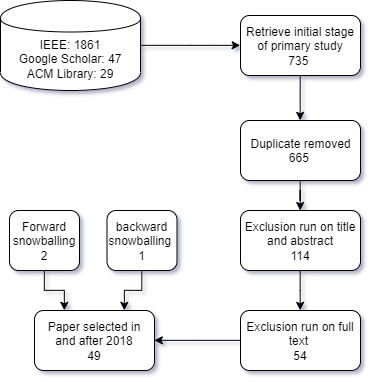}}
\caption{Attrition of papers through processing.}
\label{figure:1}
\end{figure}

Figure [\ref{figure:1}] shows the number of papers selected as the initial result and it also represents the process before selecting the number of final papers. \\

\subsection{Data analysis}
Data corresponding to the qualitative and quantitative categories was assembled in order to answer all the research questions. The papers filtered after the completion of the data extraction process, were then passed through a meta-analysis process. \\[0.5cm]

\subsubsection{Publications over time}

\begin{figure}[ht]
\centerline{\includegraphics [width=90mm]{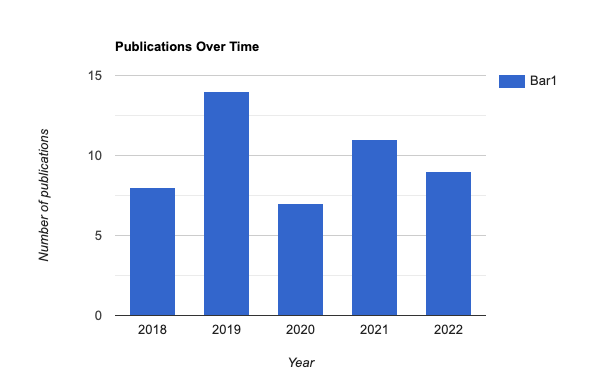}}
\caption{Number of primary studies published over time}
\label{fig:2}
\end{figure}

Considering the popularity and importance of Deep Learning technology in the real world applications, there has been significant research done in maintaining the data privacy of its models. It can be seen in Figure [\ref{fig:2}]. that the number of research published in recent years have been almost retained constantly. As we approach the future, new techniques are being introduced for privacy-preserving Deep Learning. 

\subsubsection{Significant keyword counts}
A review of keywords was performed in the primary studies in order to understand the general theme among all the chosen papers. It was noticed that apart from keywords “Deep Learning” and “privacy, the most common keywords were “Neural network”, “Parallelization”, “servers” and “Data models”.

\begin{table}[ht]
\caption{Counts of the keywords in the primary studies}
    \centering
    \begin{tabular}{p{0.5\linewidth} | p{0.4\linewidth}}
      Keyword  & Count \\ \hline
      
     Deep learning & 1842 \\
      
      Privacy & 1564\\
      Neural network  & 1424\\
      Differential privacy  & 1412\\
      Parallelization  & 1344\\
      Attacks & 765\\
      Data models  & 675\\
      Information privacy  & 580\\
      Computational modelling  & 498\\
      Distributed & 416\\
      Collaborative & 279\\
      Threats & 117\\
      
    \end{tabular}
    \label{tab:4}
\end{table}

\section{Findings} \label{sec:3}
All the primary studies were reviewed, and the analytical data was extracted from them to build Table [\ref{tab:5}]. Each primary study was based on a theme related to the framework or technique used towards achieving different levels of privacy in Deep Learning systems. The idea of every paper is noted down in Table [\ref{tab:5}]. 

Each paper’s theme has been classified further into broader categories for simplicity. Majority of primary studies can be seen focusing on the privacy-preserving models that will is the main part of this paper. Apart from that, the second r focus of our primary studies uses differential learning techniques to overcome various attacks on Deep Learning models.  

\begin{figure}[ht]
\centerline{\includegraphics [width=90mm]{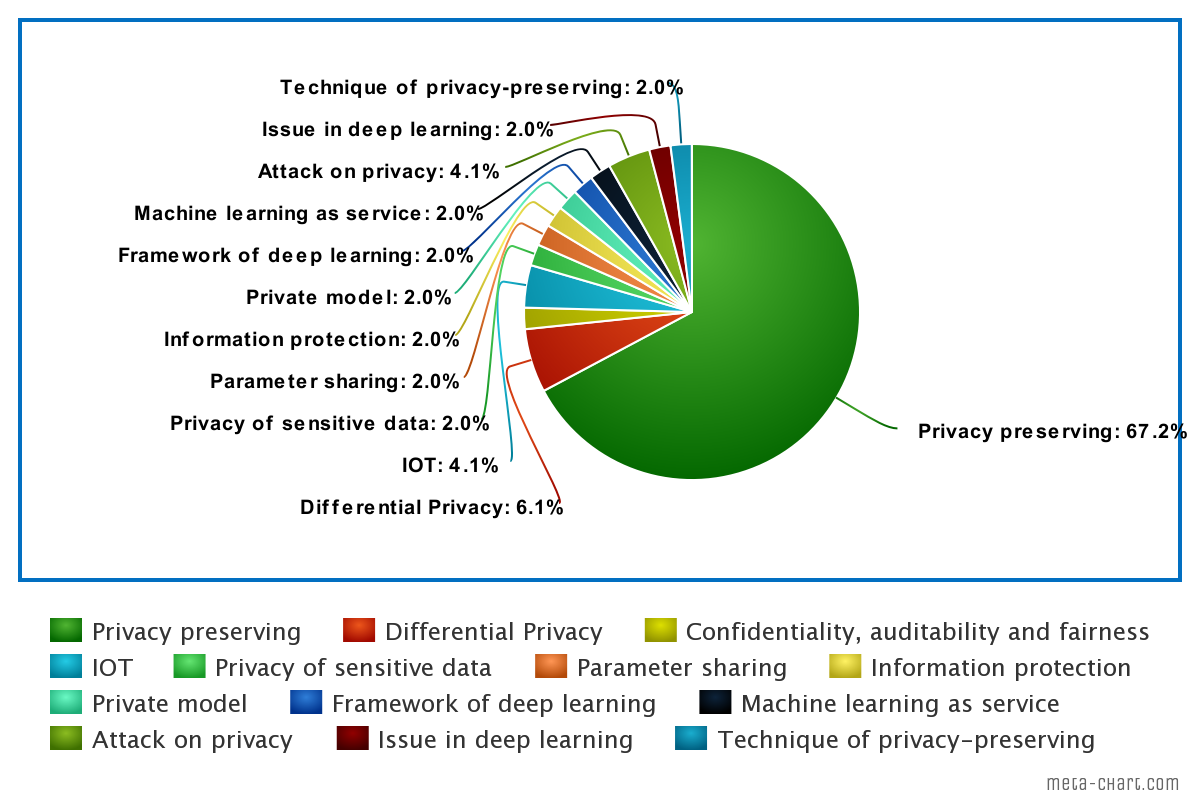}}
\caption{Chart of themes of primary studies}
\label{figure:fig3}
\end{figure}

Figure [\ref{figure:fig3}]. shows the percentage chart created after data analysis done for all the primary studies and reviewing the techniques presented in each of them.

The themes found in the primary research show that privacy-preservation is a topic that is addressed in over seventy percent (67 percent) of all works on privacy on deep learning systems. In this study, there is a mathematical definition of privacy loss that occurs to individual data records when personal data is used to create a data product, making differential privacy the second most popular subject with a proportion of 6 percent. This research primarily focuses on the ability of the device's manufacturer to monitor and abuse the data that these gadgets acquire as an issue. The third most prevalent topic, IOT, happens about 4 percent of the time. As the user adds additional devices to the system, the risk surface increases. The fourth issue is the invasion of privacy; this study contains several intrusions that threaten consumers' privacy. Nearly 2 percent of topics also include privacy-preserving techniques, deep learning problems, machine learning as a service, deep learning frameworks, private models, information protection, parameter sharing, and privacy of sensitive data.

\begin{table}[]
\caption{}
    \centering
    \begin{tabular}{p{0.13\linewidth}  p{0.53\linewidth}  p{0.22\linewidth} }
      Primary Study  & Key Qualitative and Quantitative Data
Reported & Types of Security Applications \\ \hline
      
      [\ref{ps:1}] & Differentially private learning used popularly as the go-to privacy security option in deep learning but scenarios in which it fails practically against different membership inference attacks and causes model-overfitting affect accuracy. & Privacy-preserving \\[0.5cm]

      [\ref{ps:2}] & Distributed layer-partition method which trains secured meta-data on cloud and uses stepwise activation function to provide security against privacy attacks and protects actual data. & Privacy-preserving   \\[0.5cm]

      [\ref{ps:3}] & Deep learning system with autoencoder as its core, for privacy protection in cloud based smart city applications are proposed. Homomorphic encryption and decryption techniques are used to preserve privacy. & Privacy-preserving   \\[0.5cm]

      [\ref{ps:4}] & Privacy protection distributed system for deep learning is introduced. It isolates the first layer is stored in local system while rest of the layers are deployed on centralised server and thus protecting privacy. & Privacy-preserving   \\[0.5cm]

      [\ref{ps:5}] & Introduces “DeepChain” which is a blockchain oriented distributed system for shared and secure deep training. & Confidentiality, Auditability, and fairness.    \\[0.5cm]

       [\ref{ps:6}] & The privacy of deep learning is impacted by a number of factors, including hardware performance, transmission costs, and many more. However, privacy may be achieved by combining differential privacy, homomorphic encryption, and secure multiparty computing. & Privacy-preserving   \\[0.5cm]

       [\ref{ps:7}] & The author examines the trade-off between usability and privacy in newly developed deep learning tools for genomic data-driven research. & Privacy-preserving   \\[0.5cm]

       [\ref{ps:8}] & The author's suggested adaptive clip-ping bound approach has three advances that may be summed up as follows. Author uses our particular approach to layer-wise cluster the gradient sets after demonstrating how each layer's l2 norm differs from the others. & Privacy-preserving   \\[0.5cm]

        [\ref{ps:9}] & In this study, the author used a tiered Multi-key FHE for a number of keys from the LWE assumption to maintain the privacy of datasets in the distributed deep learning. A cloud server may homomorphically calculate on ciphertexts encrypted under multiple/different participants' keys using their solution. Results may also be used to ciphertexts for extra calculation with a different key or participant.  & Privacy-preserving   \\[0.5cm]

        [\ref{ps:10}] & To protect the disclosed information, they examined the attack and defensive strategies linked to differential privacy. Taking into account the data population and deep learning architecture, they divided the attack possibilities into two categories: inference assault and system organisation.  & Privacy-preserving   \\[0.5cm]

        [\ref{ps:11}] & They research the issue of safeguarding deep learning models from MIA. The author demonstrates that standard record-DP, used to create private deep learning models, does not offer reliable and quantitative MIA protection. In addition, they suggest class-DP and subclass-DP, two novel DP ideas, as well as techniques for defending deep learning models from MIA. The class or subclass-DP can successfully fight against MIA while maintaining excellent model usefulness, according to experiments on two actual datasets. & Privacy-preserving \\[0.5cm]

    \end{tabular}
    \label{tab:5}
\end{table}

\begin{table}[]
\caption*
    \centering
    \begin{tabular}{p{0.13\linewidth}  p{0.53\linewidth}  p{0.22\linewidth} }
      Primary Study  & Key Qualitative and Quantitative Data
Reported & Types of Security Applications \\ \hline

      [\ref{ps:12}] & In this article, the author put forth two asynchronous, privacy-preserving deep learning protocols. Cooperatively, a shared model was trained in parallel and asynchronously while maintaining the privacy of both the input and the model. DeepPAR uses the proxy re-encryption approach to prevent each participant's input information from being revealed to others while maintaining the confidentiality of dynamic update information.  & Privacy-preserving   \\[0.5cm]

      [\ref{ps:13}] & A communication-efficient and privacy-preserving protocol was developed in this study by the author to allow various participants to cooperatively train a deep learning model. This protocol is useful for efficient and dependable collaborative computing among various IoT devices.  & IOT    \\[0.5cm]

      [\ref{ps:14}] & For gradient leakage robust deep learning with differential privacy, they provided a collection of algorithms with dynamic privacy settings. Using fixed-parameter techniques that inject constant differential privacy noise into all layers during each training cycle, the author first analyses various shortcomings of existing methods. The authors then provided a group of DP algorithms with a variety of dynamic parameter optimizations, such as dynamic noise scale methods, dynamic sensitivity mechanisms, and diverse combinations of dynamic parameter strategies.  & Privacy-preserving   \\[0.5cm]

      [\ref{ps:15}] & In order to overcome the major difficulties in developing a deep learning model that protects privacy, the author of this work proposes a unique technique called PDLM. In PDLM, we take into account that various DOs encrypt their data with a variety of keys before uploading it to SP. The model will then be trained by SP and CP using an effective privacy-preserving computation toolkit using the multi-key encrypted data.  & Privacy-preserving   \\[0.5cm]

       [\ref{ps:16}] & A system established on LRP algorithm in which the privacy budget is distributed dynamically to build a framework which supports differential privacy learning.  & Privacy-preserving   \\[0.5cm]

       [\ref{ps:17}] & Author introduced a new framework “SecProbe”, to address the issue of privacy-preserving collaborative deep learning systems while taking the presence of unreliable players into consideration. SecProbe uses exponential and functional mechanisms to safeguard the participants' data privacy and data quality.  & Privacy-preserving   \\[0.5cm]

       [\ref{ps:18}] & The author's suggested adaptive clip-ping bound approach has three advances that may be summed up as follows. Author uses our particular approach to layer-wise cluster the gradient sets after demonstrating how each layer's l2 norm differs from the others. & Privacy-preserving   \\[0.5cm]

       [\ref{ps:19}] & In the paper, the author discusses the privacy risks that a deep learning model faces, the fundamental idea of differential privacy, various approaches in accordance with the layer in the deep learning model where differential privacy is used. Also, presents  deep learning challenges to attain privacy protection.  & Privacy-preserving   \\[0.5cm]

    \end{tabular}
    \label{tab:6}
\end{table}

\begin{table}[]
\caption*
    \centering
    \begin{tabular}{p{0.13\linewidth}  p{0.53\linewidth}  p{0.22\linewidth} }
      Primary Study  & Key Qualitative and Quantitative Data
Reported & Types of Security Applications \\ \hline

      [\ref{ps:20}] & In this research, the author developed an approach, “PADL” to manage huge data streams in IoT applications while maintaining the privacy of training data. The framework proposed possesses the property of asynchronous optimization and the privacy property against severe collusion attack.   & IOT    \\[0.5cm]

      [\ref{ps:21}] & The author of this paper describes our deep-learning-based service provision system for delivering increased services and privacy protection in edge computing. The authors' method for privacy protection involves performing the private dense training phase and the private compressive training step, where the dense model and compressive model, respectively, are protected through differential privacy.   & Privacy-preserving    \\[0.5cm]

      [\ref{ps:22}] & In this study, the author suggests tackling the issue of the privacy of sensitive healthcare data while utilising DL algorithms by using the PHE-based Paillier method.  & Privacy-preserving   \\[0.5cm]

      [\ref{ps:23}] & They suggested a technique for exchanging models sequentially throughout the distributed neural network training process. They presented approach might be used because of its own privacy-preserving nature, even when the training datasets are gathered by many agencies but cannot be shared by other agencies owing to some legal and ethical problems. Privacy-preserving distributed deep learning training makes use of the parameters of the training model rather than the training dataset.  & Privacy-preserving   \\[0.5cm]

       [\ref{ps:24}] & Here, the author analysed recently created information protection techniques and evaluated the GAN model attack in-depth and methodically. The key benefit of the suggested solution is that it is more precise and flexible than the information protection techniques now in use. However, there are several places where the GAN model attack may be improved upon. & Information Protection   \\[0.5cm]

       [\ref{ps:25}] & With three unique contributions, they introduced our method of differentially private deep learning for model publication. First, they use CDP for privacy accounting to get precise assessment of privacy loss because training neural networks requires a lot of iterations. Second, they separate two distinct data batching techniques and provide privacy accounting techniques for each technique's calculation of privacy loss.   & Private Model   \\[0.5cm]

       [\ref{ps:26}] & To train a deep neural network with great privacy and high accuracy, the author proposed a novel LDP technique. Compared to other differentially private techniques, it exhibits outstanding accuracy even at very low privacy budgets. The proposed framework uses untrusted curator setting which offers a better level of privacy while reducing computational work, in contrast to the existing differentially private systems require a trusted curator since they use global DP to implement them.  & Privacy-preserving   \\[0.5cm]

    \end{tabular}
    \label{tab:7}
\end{table}

\begin{table}[]
\caption*
    \centering
    \begin{tabular}{p{0.13\linewidth}  p{0.53\linewidth}  p{0.22\linewidth} }
      Primary Study  & Key Qualitative and Quantitative Data
Reported & Types of Security Applications \\ \hline

      [\ref{ps:27}] & In this study, the author explored privacy concerns in deep learning and compared several privacy-preserving methods to counter these dangers, including homomorphic encryption, differential privacy, garbled circuits, etc. Analysed these various methods in-depth and provided an overview of how well existing solutions performed. Talked about some unresolved issues and difficulties in this field.    & Privacy-preserving    \\[0.5cm]

      [\ref{ps:28}] & In this work, the author proposed "Privacy Partition" as a system for limiting an adversary's ability to undertake input recovery attacks if they have access to an intermediate activation or significant piece of a deep network topology. A deep network privacy partition makes the local layer operations more invertible, which reduces the likelihood that network inputs may be recovered from intermediate network states.  & Privacy-preserving    \\[0.5cm]

      [\ref{ps:29}] & By taking advantage of the privacy flaws in the stochastic gradient descent process, the author in this study designed and assessed novel white-box membership inference attacks against neural network models. In both centralised and federated environments, in context of both passive and active inference attackers, and assuming various adversary prior information, the study illustrated the proposed attacks. The research also demonstrated that such white-box membership inference attacks significantly affects even well-generalized models.   & Privacy-preserving   \\[0.5cm]

      [\ref{ps:30}] & In this research, the author introduces “ATP”, an anonymous distributed deep learning protocol that adds temporary random noise to a large synthetic mini-batch size of sharing gradient approach. These values were then transmitted to a private network for synthesis and interference removal. Even in cases where the fusion server colluded with additional n-2 participants, the suggested architecture enables safety of the supplied gradients.   & Privacy-preserving   \\[0.5cm]

       [\ref{ps:31}] & This paper introduced a “PDLHR” for use with multisource robot systems that addresses privacy-preserving and multikey issues, assures the safety of collaborative training of multisource robot data, increases training effectiveness, and minimises interaction while maintaining privacy. Suggested approach offers a theoretical foundation for multisource collaborative robots' ciphertext deep learning training as well as a foundation for robot systems in smart grid or other robot situations.  & Privacy-preserving   \\[0.5cm]

       [\ref{ps:32}] & In this article, the author addressed some of the difficulties in applying differential privacy to machine learning, including selecting the best parameters and how they affect the outcomes, determining the best balance between accuracy and privacy loss, and the drawbacks and advantages of the available differential privacy machine learning libraries. The findings indicated that the most crucial phases are selecting the appropriate privacy settings and DNN architecture parameters.    & Privacy-preserving   \\[0.5cm]

    \end{tabular}
    \label{tab:8}
\end{table}

\begin{table}[]
\caption*
    \centering
    \begin{tabular}{p{0.13\linewidth}  p{0.53\linewidth}  p{0.22\linewidth} }
      Primary Study  & Key Qualitative and Quantitative Data
Reported & Types of Security Applications \\ \hline

      [\ref{ps:33}] & This study provided a collaborative method for training CNN models with diverse datasets from various sources while maintaining privacy. In this approach, CNN models were trained in a distributed learning environment without sharing or combining the datasets beforehand. First, only the trained CNN network is shared with others after each dataset has been utilised to train the network independently in its own domain. The trained network is then put to the test using test photos across several domains.     & Privacy-preserving    \\[0.5cm]

      [\ref{ps:34}] & This study developed a generative adversarial microaggregation method that protects the privacy of IoT data. To produce realistic samples based on their estimated distribution, the technique learns the original data distribution. On several classifiers, the authors evaluated the resulting datasets' accuracy and privacy. The outcomes showed that the privacy-utility trade-off was competitive with already available methods.   & Privacy-preserving    \\[0.5cm]

      [\ref{ps:35}] & A technique for using crowd-sourced data and computing resources to train a federated learning model using blockchain has been described. Authors have presented an IPFS and RBAC smart contract-based blockchain-based data storage and sharing system that also communicates the training data with PRE. For the crowd-sourced model training on the proposed hybrid blockchain architecture, a novel APBFL algorithm has been introduced. In order to test the success of suggested technique, a TextCNN model was trained in the simulated network. Extensive numerical results supported the method's efficacy in terms of efficiency, security and robustness.     & Privacy-preserving   \\[0.5cm]

      [\ref{ps:36}] & This study presented “DeepSub”, a novel framework for subset selection, it demonstrated its value in choosing an exemplary subset of training examples for training deep neural networks. Applications with severe resource limitations, where it is impractical to use GPUs and other processing resources, greatly benefit from this framework. The system showed great promise in selecting an informative training subset to train a range of deep learning models on three difficult computer vision applications.    & Framework of deep learning   \\[0.5cm]

       [\ref{ps:37}] & The study introduced a decentralised machine learning approach that, depending on input perturbation, ensures differentiated privacy. The comparative findings demonstrated that the suggested strategy accomplishes learning more quickly and accurately while maintaining the necessary level of anonymity and privacy. It was verified that while adding more nodes speeds up learning, adding too many nodes worsens runtime characteristics because of increased communication delay.   & Privacy-preserving   \\[0.5cm]

    \end{tabular}
    \label{tab:9}
\end{table}

\begin{table}[]
\caption*
    \centering
    \begin{tabular}{p{0.13\linewidth}  p{0.53\linewidth}  p{0.22\linewidth} }
      Primary Study  & Key Qualitative and Quantitative Data
Reported & Types of Security Applications \\ \hline

      [\ref{ps:38}] & This study discusses different cryptographic primitives which have been employed depending on the privacy-preserving purpose and situations. These include Homomorphic Encryption, Garbled Circuits and Goldreich-Micali-Wigderson protocols. They also included secret sharing. Although each of these approaches has a number of benefits, none of them fully satisfies the requirements for the many privacy-preserving applications that are of relevance. In order to offer the greatest qualities to each individual application, a number of recent proprietary machine learning approaches are based around a hybrid mix of the fundamental protocols. This paper discussed both secure basic and hybrid protocols.      & Privacy-preserving    \\[0.5cm]

      [\ref{ps:39}] & They provide a thorough analysis of the most recent PPDL on MLaaS in this study. They explore both using DL for PP as well as the traditional PP approach. Their study also tackles the difficulty of using cutting-edge DL methods with PP, and they analyse the NN's original structure and the adjustments required to utilise it in privacy-preserving settings. They also suggested a multi-scheme PPDL classification based on an adversarial model, PP approaches, and the difficulties and shortcomings of current PPDL techniques.    & Machine learning as service     \\[0.5cm]

      [\ref{ps:40}] & This study analysed the degree of privacy guarantee and the resilience of differentially private GAN models to the membership inference attack. It was depicted in the experimental assessment that differential privacy may lower the attack success rates of membership inference while maintaining the quality of synthetic data by evaluating the efficacy of the attack depending on the level of privacy guarantee. We discovered that loosening the concept of differential privacy comes with extra privacy issues after looking into numerous theories of differential privacy.      & Privacy-preserving   \\[0.5cm]

      [\ref{ps:41}] & A differentially private and quick technique of protection against membership inference and model inversion attacks was put forth in this research. In order to confound the attacker's model, they employ an exponential technique to change and normalise the confidence score vectors. & Attack on Privacy   \\[0.5cm]

       [\ref{ps:42}] & Their threefold primary contribution is to this paper. In the beginning, they suggest a differentially private deep learning algorithm that, when compared to earlier approaches, leads to a faster convergence and higher accuracy. Second, they quantitatively demonstrate that ADADP meets differential privacy using more sophisticated analytical techniques after intuitively analysing the advantages of ADADP versus DPSGD.    & Differential Privacy    \\[0.5cm]

    \end{tabular}
    \label{tab:10}
\end{table}

\begin{table}[]
\caption*
    \centering
    \begin{tabular}{p{0.13\linewidth}  p{0.53\linewidth}  p{0.22\linewidth} }
      Primary Study  & Key Qualitative and Quantitative Data
Reported & Types of Security Applications \\ \hline

      [\ref{ps:43}] & We suggest the adaptive differentially private deep learning model in this work. According to intuition, we first clip the gradient to limit the sensitivity, then we inject differentially private noise into the clipped gradient with a certain decay rate based on the Gaussian process, and last we update the gradient with SGD.       & Privacy-preserving    \\[0.5cm]

      [\ref{ps:44}] & They use the adversarial example concept to suggest a framework to safeguard image privacy. The two contributions made by this study are substantial. They begin by defining two new measures for image privacy. In order to maximise the suggested two metrics, they next build two privacy protection systems.    & Attack on privacy      \\[0.5cm]

      [\ref{ps:45}] & For broad deep learning problems, they suggest a system in this research that offers differentially private prediction probability vectors. their method only introduces DP noise into a single neuron at the network's output layer.  & Differential Privacy   \\[0.5cm]

      [\ref{ps:46}] & This research suggests a brand-new approach for protecting deep learning privacy. First off, this article employs the LBP approach to extract data information rather than the more common convolution feature extraction, which significantly decreases the information extraction dimension.      & Privacy-preserving   \\[0.5cm]

       [\ref{ps:47}] & They provide a thorough overview of differential privacy and its uses in this work. In deep learning, federated learning, and data collecting, differential privacy and local differential privacy ensure significant privacy protection for consumers' personal information.     & Differential Privacy    \\[0.5cm]

       [\ref{ps:48}] & In this work, the authors first discuss the possible dangers of DL before reviewing the two types of attacks that can occur—model extraction assault and model inversion attack—as well as the four standard protection technologies—DP, HE, SMC, and TEE—that are used to safeguard the privacy of user data. Then they looked at two attack types: poisoning attacks and adversarial attacks.      & Issue in deep learning     \\[0.5cm]

       [\ref{ps:49}] & The current situation of PPDL has been addressed in this document. We examine the basic structure of the neural network and the modifications required for its application in a setting that protects privacy. We also address the major issue, which is the trade-off between accuracy and complexity during the process of substituting non-linear activation function. Reducing computational load is an unsolved issue with privacy-preserving machine learning techniques.      & Technique of privacy-preserving      \\[0.5cm]

    \end{tabular}
    \label{tab:11}
\end{table}

\section{Discussion} \label{sec:4}

The chosen primary studies unmistakably show how crucial deep learning privacy is. The majority of the efforts are based on differential learning, then privacy-preserving deep learning. There is a tremendous amount of data and information created every minute in this age of technology, thus it is crucial to strive toward protecting the privacy of this data and how we choose to retain and use it. However, a distributed deep learning approach, which has lately drawn a lot of interest, can overcome this problem. The privacy problem sadly gets worse when comparing a distributed deep learning situation to a conventional solo deep learning scenario.

The researchers provide a distributed layer-partitioned training strategy that protects data privacy while allowing metadata (not the original data) to leave a local site. The main concept is that before leaving a local site, the original data is transformed into irreversible secured metadata. With the help of this technology, significant training may be done in a distributed manner on the permanently secured metadata of distant sites while the original data is kept at the local permanently secured site [\ref{ps:3}]. 

The training strategies proposed in most of the studies to preserve privacy adds noise to the data and increase the budget of the system. The adaptive allocation dynamic privacy budget differential privacy learning (ADDP) framework is suggested in the paper [\ref{ps:16}] and is based on Layer-wise Relevance Propagation (LRP). First, while utilizing the LRP to calculate the relevance, add noise to the relevance decomposition data. Second, the size of the privacy budget is constantly altered during the training process, and this privacy budget is adaptively assigned according to the relevance, in order to add noise to the gradient.

An increasing interest in the field of IoT is also witnessed in some of the primary studies as it has recently seen significant development and has become more crucial to our daily lives. The fact that the Internet of Things will produce a lot of data is one of its greatest advantages. These vast amounts of data cause various techniques, particularly deep learning, to improve quickly. Deep learning has a significant potential to uncover great values from the massive amounts of data generated by IoT devices and is capable of learning complicated properties. 

As mentioned earlier, in order to protect privacy, past research often adds differentially private noise to all neural network parameters. As a resolution, [\ref{ps:8}] offer a unique approximation mechanism for obtaining a polynomial-form loss function and then directly injecting differentially private noise into the coefficients of the approximated loss function. By injecting noise into only a portion of the parameters, the amount of noise introduced is considerably reduced while maintaining the privacy level. Because each participant's noise is decreased, the total noise in large-scale distributed learning is lowered dramatically, and performance is greatly improved. Another way to preserve privacy of IoT based Deep Learning models is proposed in [\ref{ps:20}] where the researchers have proposed a viable approach i.e., privacy-aware and asynchronous deep-learning assisted IoT applications (PADL) for IoT applications to ensure training data privacy while handling huge data streams in IoT applications. The PADL has the property of asynchronous optimization and the privacy property against the extreme collusion attack, according to the properties study.

Talking about applying Differential Learning to preserve privacy, [\ref{ps:45}] and [\ref{ps:47}] studies present the frameworks that could be used along with the Deep Learning models. Differential privacy (DP) [\ref{ps:5}, \ref{a10}] is a promising technique that (informally) conceals the existence of any random data sample in a dataset, preserving membership privacy for all members of the training set. The techniques proposed in the primary studies mainly instruct adding noise to the data but as an improvement present various ideas to decrease noise and keep quality data in place maintaining security at the same time.\\[0.5cm]

\textbf{RQ 1. What are the different types of privacy attacks on a deep learning system?} 

Deep learning systems are being deployed everywhere around us, however, such widespread use makes them susceptible to several types of attacks. In most cases, the data feed to the deep learning architectures is highly sensitive and confidential and requires protection and security during the complete life cycle.

One such popular attack is membership inference attacks, which may reveal sensitive information about a person's participation in a training data set [\ref{ps:1}]. These types of attacks pose a big threat to differential privacy scheme deployed in deep learning [\ref{ps:1}]. 

Another class of attacks are called “Model Extraction Attacks” [[\ref{ps:48}], one such attack was presented by Tramer et al. [[\ref{rs:10}] with the goal of copying the parameters of ML models used to deliver cloud-based ML services. The fundamental concept is to construct model equations from the results produced by submitting several queries. Another attack under this classification aims to steal the hyperparameter of the target model, Wang et al. [\ref{rs:12}] suggested hyperparameter theft attacks.

“Model Inversion Attacks” [\ref{ps:48}] are class of attacks in which the adversary seeks to reveal the secrecy of private records that were utilised as part of the training set by using model predictions. 

Tracing (membership inference) attacks and reconstruction attacks are the two primary forms of inference attacks in deep learning [\ref{rs:11}]. In reconstruction attacks, the attacker wants to take training data out of the model output forecasts. Another attack is “Man in the middle attack” Attacks that place the victim in the centre can be divided into Active Attacks and Passive Attacks. The passive attack listens in on data exchange between two devices passively. Even though the passive assault violates privacy, they leave the data unchanged. With access to the device, an attacker can discreetly observe for a considerable amount of time before launching the attack and gathering all the data [\ref{ps:48}].

Moreover, there are several other attacks on deep learning system such as “Data Inference attack” in which a membership inference or a model inversion assault are two types of data inference attacks, each of which has a distinct inference objective. Both techniques provide attackers access to the target model, which is reviewed by its owner [\ref{ps:44}]. Another one is “Differential Privacy attacks” In which, an individual’s record is added to or removed from a dataset. T thing has little impact on the dataset's analytical results. Artificial intelligence, multi-agent systems, and cyber security are just a few of the study fields where differential privacy has been widely used [\ref{ps:44}]. “Membership Interface Attacks” which target broad machine learning models and sought to use the confidence vectors provided from the target model to infer if certain data were included in the training dataset [ \ref{ps:40}]. “White Box attack” in which, we assume that in the ideal white-box situation, the attacker has access to the trained GAN model's discriminator and uses it as the attack engine (or attacker model) for membership inference [\ref{ps:33}]. \\[0.5cm]

\textbf{RQ2: What are the various privacy-preserving solutions for the different attacks? }

When the model parameters are revealed, the training data can be secured from inversion or inference attacks by using differential privacy in deep learning models. Some of our primary studies demonstrates the use differential privacy to deep learning models.

Studies [\ref{ps:2}] and [\ref{ps:11}] provide security against model inversion attacks. [\ref{ps:3}] proposes “DeepChain”, a blockchain foundation distributed system for shared and secure deep training ensuring confidentiality, auditability and fairness and providing security against inference attacks. [\ref{ps:4}] introduces another distributed deep learning system dividing the deep architecture layers among local and centralised servers to ensure privacy and security of confidential data.

Employs homomorphic encryption and decryption in the user module and runs the computations on the cloud, to maintain privacy [\ref{ps:5}]. Deep learning classification uses an autoencoder, and the outcomes are evaluated using a variety of performance parameters, including accuracy, number of cloud nodes, training time, and encryption time.

[\ref{ps:7}] performs exceptionally against membership inference attack. [\ref{ps:9}] proposes an LWE-Based Multi-Key approach to protect against collusion attacks.[\ref{ps:13}] provides a system which is resistant to generative adversarial network attacks. Whereas [\ref{ps:14}] provides a system which is protected from gradient leakage attack.

Some techniques employ the F1 score and the precision of the membership inference attack as assessment indices. Reiter et al. [\ref{rs:24}] assessment of the privacy risk of data publication by inference on synthetic data sets was impractical since it would need running a series of inference assaults. It will be a desirable security solution for deep learning applications if a model designer can show that differential privacy is effectively protected on both theoretical proof and experimental outcomes [\ref{ps:19}]. 

Moreover, to safeguard data privacy, a federated learning architecture was created for IoT Edge devices. Federated learning is nevertheless accomplished at the cost of a large communication overhead [\ref{ps:20}]. 

Data security requires encryption, and sluggish connections are recommended to reduce packet loss and prevent data loss [\ref{ps:22}].Protection from attacks and privacy risks. To reduce the privacy risk associated with output model parameters, differentially private deep learning tries to calculate model parameters in a differentially private manner. Model inversion attacks and membership inference attacks are two examples of well-known deep learning attacks. Attacks using model inversion use model access and prediction output to deduce input instances. Attacks using membership inference take use of the prediction API's black box access to determine the membership of certain training instances [\ref{ps:25}].

Data-oriented attacks change a program's innocuous behaviour without endangering the integrity of its control-flow by manipulating non-control data. It has been demonstrated that even in the presence of control-flow defensive systems, such assaults are still capable of causing severe harm. These dangers haven't, however, been sufficiently addressed. Data-oriented exploits, such as Data-Oriented Programming (DOP) assaults, are first mapped to their assumptions/requirements and attack capabilities in this SoK study [\ref{ps:29}]. If an attacker is introduced into the consensus group and attempt to destroy the final model, the situation drastically changes. Training procedure using a standard federate technique without cross-verification and Node 0 delivering a random model as opposed to a trained one as the baseline[\ref{ps:35}]. Comparing different ideas of differential privacy to the non-private situation, differential privacy can dramatically reduce privacy leakage. It would be interesting to research privacy leaking on differentially private algorithms in a future work using tactical techniques since this would help with attack defence [\ref{ps:40}].  

Depending on the strategies used, defence techniques may be loosely divided into three categories: 1) Regulation-Based Defense Techniques: The first strategy involves limiting the prediction vector to the top k classes; a lower k results in less information leakage. The second technique involves rounding the classification probabilities included inside the vector down to d floating point digits in order to reduce the accuracy of the confidence score vector. Once more, a lower d indicates less information leakage. 2) Examples-Based Adversarial Defense Techniques: Their theory is based on the observation that hostile instances can cause deep learning algorithms to make incorrect predictions. 3)Deep-Network-Based Defense Techniques: One of the five locations in a deep neural network where noise can be inserted to accomplish differentially private deep learning is: source datasets gradient loss functions neural network weights and output classes [\ref{ps:42}]

\textbf{RQ3: What are the methods by which we can preserve the privacy of users and reduce the communication burden?}

Reducing communication barriers to reduce noise and enhance model performance, the study has offered a novel privacy-preserving technique [\ref{ps:8}]. Additionally, one has drastically decreased the upload and download parameters to save the cost of communication [\ref{ps:8}]. Three authentic data sets were used in our experiment, which involved two different neural network topologies [\ref{ps:8}].

A range of 30 to 150 devices were examined. According to the experimental findings, we may save 98.5 percent on communication costs as compared to complete gradient exchanges. One can improve accuracy by up to 16 percent over the prior work (=1, CNN model for SVHN). Approximate mechanism technique may be immediately extended to other deep learning models (such the recurrent neural network), which points to our future direction even if this study concentrates on MLP and CNN [\ref{ps:8}].

Adapting the system to the recently developed federated learning scenario is another fruitful area for IoT application development[\ref{ps:8}]. Its realized that malevolent adversaries conducting security assaults, such as a poisoning attack or an inference attack, will jeopardize the integrity and availability of the scheme. Nevertheless, it is continued to assert that our strategy rigorously ensures the privacy of the training data and that the local model theoretically meets -differential privacy [\ref{ps:20}].

\section{Future research directions of deep learning privacy} \label{sec:5}

We suggest the following research directions for privacy of Deep Learning systems that merit additional study based on the findings of this survey and our review of primary  studies:\\[0.5cm]

\textbf{Differential privacy in deep learning:} Gradient-level approach is one way to implement Differential privacy. The gradient-level method has two flaws. To begin, this technique must entirely trust the parameter server. If the attacker is a parameter server, the victim's private information is revealed by collecting and reversing the gradient descents of the parameter server's members in the collaboration model [\ref{rs:13}], [\ref{rs:14}]. Second, participants of the collaborative learning model send the same model as the server. This is a problem that results in one member knowing the model architecture of other members. In this scenario, the adversary is a collaborative learning participant who can use a white-box strategy to steal information from other participants [\ref{rs:15}]. However, we can use a separate privacy protection strategy to each member in a learning model, and the parameter server can aggregate the members' learning model contribution while still protecting the private of the member's model.

\textbf{Deep learning privacy in IoT:} The IoT-related research resulted in the development of a communication efficient and privacy-preserving protocol to allow different participants to collaboratively train a deep learning model, which is of general value for efficient and reliable collaborative computing among different IoT devices. To reduce noise and increase model performance, a novel privacy-preserving technique is presented. Furthermore, both upload and download parameters have been decreased in order to considerably cut communication costs. The trial findings indicated that we may save 98.5 percent of the overall cost of communication when compared to full gradients exchanges. The proposed mechanism can be immediately extended to other deep learning models (for example, the recurrent neural network), pointing the way forward. Extending this approach to the rapidly developing federated learning scenario is another intriguing research path in IoT applications.

Differential privacy is now mostly utilized to address the privacy security issue of deep learning training data, and it only provides a limited defence against membership inference attacks. In fact, while the challenges to deep learning in practical applications are diverse and complicated, they all share the goal of over-fitting deep learning.

Over-fitting is one cause of privacy leakage in machine learning models, but it is not the only one. It is an intrinsic difficulty in machine learning that restricts the model's prediction accuracy and generalization capabilities. There is evidence that differential privacy can even avoid over-fitting to minimize prediction errors in very large data sets [\ref{rs:16}]. It suggests that machine learning and privacy researchers may not necessarily be competing in a zero-sum game, but rather share comparable interests. Lécuyer et al. [\ref{rs:17}] pointed out that differential privacy may be utilized to defend against hostile cases [\ref{rs:18}], which broadens differential privacy's application area.

\section{Conclusion and future work} \label{sec:6}

This paper is based on the latest research done in the field of Deep Learning systems and its privacy-preserving models. We reviewed and discussed various frameworks, privacy preserving models and solutions for possible network attacks on Deep Learning systems to maintain the confidentiality, integrity and authenticity of the sensitive information and important data. The vast applications of Deep Learning involves Healthcare, Economic areas, Finance and Industry. We have tried to cover the privacy concerns in almost every field. The broad usage of Deep Learning encourages researchers to enhance the security and privacy of its models and to deter the malicious activity related to these models.

DL models are quite complex. As a result, creating a new model is difficult. After developing a DL model, several users utilize the model and train it using their own data, such as DNN for image classification. If these models are used in public, they may be vulnerable to an attack such as a black-box. As a result, we must exercise caution while implementing the approach, particularly on large scale variations [\ref{rs:19}], [\ref{rs:20}]. However, preserving data privacy remains the most significant issue. As a result, developing safe DL models for customized data while maintaining data privacy is an intriguing direction to take. Furthermore, further effort is required to develop techniques to prevent these attacks not just in the literature but also in the real-world setting.

\textbf{Potential research agenda 1:} Recent research has offered numerous privacy-preserving DL techniques to safeguard the privacy of sensitive data. However, there is still a lot of work to be done before it can be used in practice. The most significant impediment to privacy-preserving DL approaches is computing expense. The calculation cost of DL is considerable because to its non-linear behaviour, which severely limits its availability. One significant difficulty for privacy-preserving DL approaches is to reduce the overhead of privacy-preserving DL techniques.

\textbf{Potential research agenda 2:}  A high-performing ML model needs vast volumes of training data, extensive hardware resources, and a significant amount of time for parameter adjustment. As a result, the labelled training dataset, model architecture, and model parameters have been identified as commercial intellectual property that must be safeguarded. There are currently just a few works on intellectual property protection for watermark-based machine learning models [\ref{rs:21}], [\ref{rs:22}], and [\ref{rs:23}], and their efficacy is challenging to insure. A more effective and safe intellectual property protection mechanism for neural network models has yet to be developed.

\textbf{Potential research agenda 3:} Most existing privacy-protection systems can only predict privacy during the testing phase, and only a few solutions can be trained on encrypted data. Furthermore, unencrypted data is used to train the inference privacy-preserving models, and the trained weights and biases are then applied to alternative models in which the activation function is substituted with a basic activation function, such as a square function. Differences between the trained and inferred models often result in a significant decrease in model performance. As a result, existing privacy-preserving approaches still need a significant amount of customization for each DL model. A broad framework that protects privacy is a problem that must be solved in the future.\\ [0.5cm]
\textbf{Declarations of interest}

None\\ [0.5cm]
\textbf{Acknowledgement}\\

we would like to express our heartfelt gratitude to professor Ali Dehghantanha for their assistance in completing the project.

\end{document}